\documentclass[10 pt,final,journal,letterpaper,oneside,onecolumn]{IEEEtran}
\usepackage{multicol}   
\usepackage{multirow}
\usepackage[pdftex]{graphicx}
\usepackage[lined,boxruled]{algorithm2e}
\usepackage{algorithmic}
\usepackage[small]{caption}
\usepackage{float}
\usepackage{xspace}
\usepackage{fancyhdr}
\usepackage{amssymb,amsmath}
\usepackage{subfig}
\usepackage{textcomp} 
\usepackage{epstopdf}
\usepackage{bbding}
\usepackage{color}
\begin{document}

\title{Internet of Autonomous Vehicles: Architecture, Features, and Socio-Technological Challenges}

\author{Furqan Jameel, Zheng Chang, Jun Huang, Tapani Ristaniemi}

\markboth{Accepted in IEEE Wireless Communications}%
{\MakeLowercase{\textit{et al.}}: Wireless Communications}

\maketitle

\begin{abstract}
Mobility is the backbone of urban life and a vital economic factor in the development of the world. Rapid urbanization and the growth of mega-cities is bringing dramatic changes in the capabilities of vehicles. Innovative solutions like autonomy, electrification, and connectivity are on the horizon. How, then, we can provide ubiquitous connectivity to the legacy and autonomous vehicles? This paper seeks to answer this question by combining recent leaps of innovation in network virtualization with remarkable feats of wireless communications. To do so, this paper proposes a novel paradigm called the Internet of autonomous vehicles (IoAV). We begin painting the picture of IoAV by discussing the salient features, and applications of IoAV which is followed by a detailed discussion on the key enabling technologies. Next, we describe the proposed layered architecture of IoAV and uncover some critical functions of each layer. \textcolor{black}{This is followed by the performance evaluation of IoAV which shows significant advantage of the proposed architecture in terms of transmission time and energy consumption.} Finally, to best capture the benefits of IoAV, we enumerate some social and technological challenges and explain how some unresolved issues can disrupt the widespread use of autonomous vehicles in the future.    
\end{abstract}


\IEEEpeerreviewmaketitle

\section{Introduction}
\begin{figure*}[!htp]
\centering
\includegraphics[trim={0 0cm 0 0cm},clip,scale=.9]{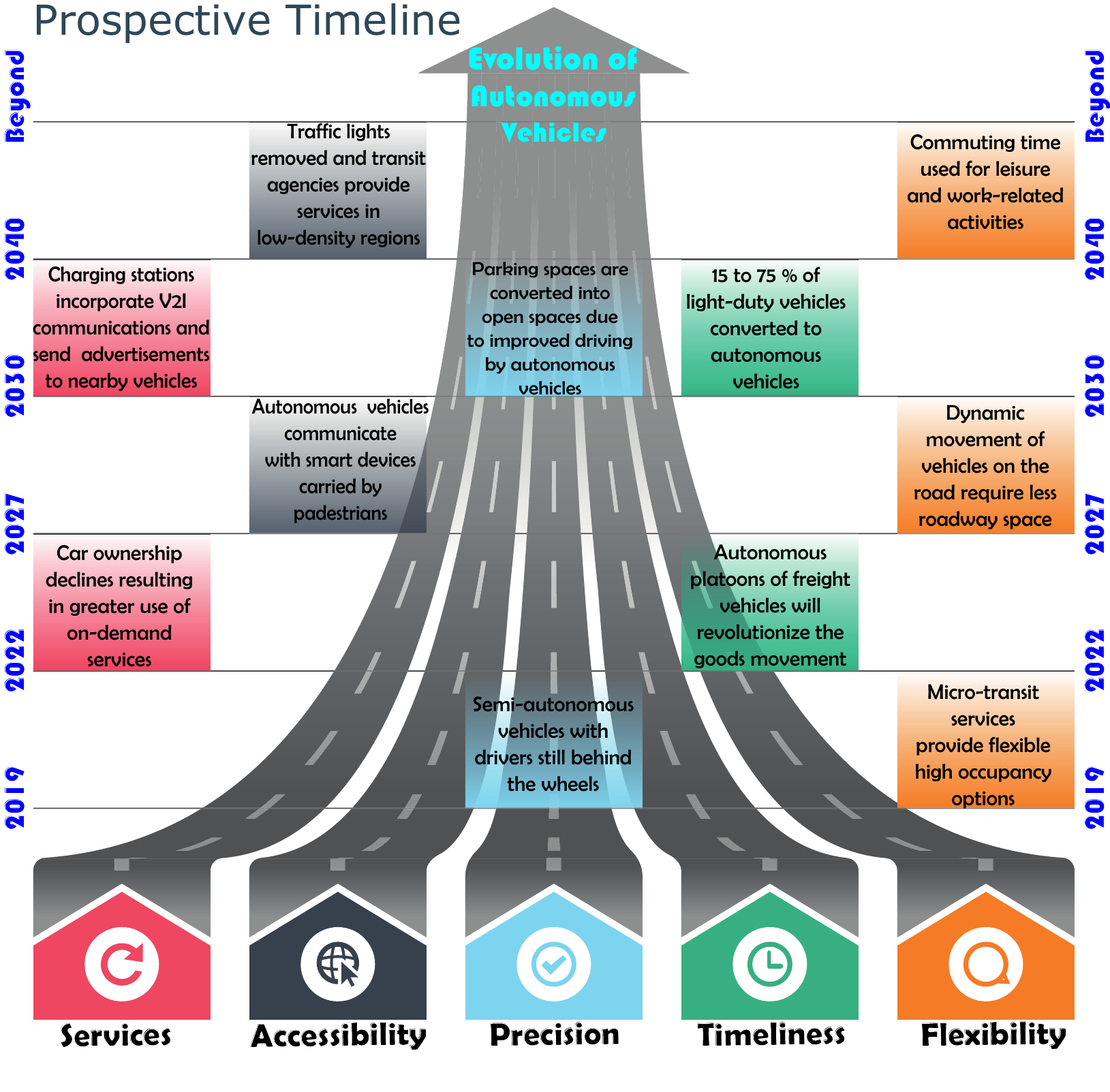}
\caption{\textcolor{black}{Prospective time-line of evolution of autonomous vehicles.}}
\label{fig.1}
\end{figure*}

The development of the automotive landscape in the coming years will be significantly influenced by the global political and economic conditions. The idea of sustainability against the background of dwindling and rapidly increasing supplies of fossil fuels will also be consolidated on a global scale \cite{hannon2016integrated}. The desire to own a car will be significantly reduced in developed countries in favor of integrated mobility solutions \cite{litman2017autonomous}. \textcolor{black}{Subsequently, there is going to be a lasting change in the transportation technologies used for commuting and logistics. The electrification of vehicles, as well as the use of alternative fuels, will lead to an increase in the number of charging stations and bring extremely high diversity in vehicular technology \cite{zhang2018vehicular}. In the process, as shown in Figure \ref{fig.1}, semi- and fully-autonomous vehicle are expected to grow gradually and vehicular communications aspects like accessibility, timeliness, and flexibility will continue to improve in coming decades.} Therefore, we anticipate that the next evolutionary step in vehicular communications is going to be the Internet of autonomous vehicles (IoAV).

Air traffic is often used in the context of autonomous connected vehicles, in which - apart from takeoff and landing - the autopilot takes over the control. However, there are several differences between road traffic and air traffic (including the type of transport), which does not make a transfer between the systems meaningful and feasible. One of the key differences is the application of rules and the type of control. In areas where aircraft meet each other aircrafts - especially near runway- everything is determined by strict rules. Compared to air traffic, road traffic is more of a self-organized chaotic system and arranged principally by flexible rules \cite{litman2017autonomous}. \textcolor{black}{Moreover, despite carefully regulating the traffic laws, many random unanticipated situations cannot be defined as a clear rule.} Participation on the road requires constant caution and mutual consideration in a variety of circumstance. Thus, one of the central tasks of a driver for safe participation in road traffic is the assessment of the behavior of other road users. This assessment includes communicating with the drivers and road users through actions and signs \cite{zhang2018vehicular}. In other words, besides the \emph{official} rules, there is a set of informal rules that control traffic. From a systemic point of view, it is immediately obvious that observance of informal rules, communication between road users, and behavioral prediction of drivers are important to support the flow of traffic and to minimize the number of accidents on the road \cite{joy2018internet}.

\textcolor{black}{The concepts of autonomous vehicles was introduced in the late 20th century. At that time, the misconduct of motorists was identified as the main cause of the accident.} The fact that infrastructure and vehicle design are also decisive factors in the number and severity of the accident, was not seriously considered. Later on, the idea of a machine-based substitution of the human driver was surfaced which led to technologies like cruise control and self-driving vehicles. Today, the developments in embedded systems and wireless communications have changed the requirements and the industry giants are competing to claim the top spot in autonomous vehicle technology. \textcolor{black}{Millions of dollars are being invested to improve different aspects of autonomous vehicles. Intel, Verizon and Samsung are some of the leading investors with projects like INRIX, Telogis, and Joynet, respectively \cite{hannon2016integrated}. These project cover a wide range of aspect including predictive traffic information, Cloud-based GPS tracking and efficient 3D mapping.} 

Although the upsurge in data rate requirements has paved the way to the Internet of vehicles (IoV), the autonomous part of IoV is still not fully explored \cite{gupta2018authorization}. \textcolor{black}{The future vehicles will be a formidable sensor platform and they will absorb information not only from the environment but also from other vehicles (drivers). Subsequently, this large amount of information will be fed to the infrastructure. A comparison of IoV and IoAV is provided in Table \ref{tab_1}.} Adopting a different approach from existing works, where technical prototypes, literary metaphors, and pictorial imaginations are clashed but never synchronized, the fundamental objective of this paper is to provide a functional ecosystem for the legacy and autonomous vehicles. More specifically, we aim to create a layered transport communication fabric capable of safe navigation and efficient traffic management while quickly driving people to their destinations. \textcolor{black}{We will also identify the key features of this ecosystem and see how IoAV can provide performance improvements through the layered architecture. Finally, we discuss the simulation results for performance evaluation of IoAV and discuss the socio-technological challenges for future research.} 

\begin{table}[]
\centering
\caption{\textcolor{black}{Classification of different attributes of IoV and IoAV.}}
\label{tab_1}
\textcolor{black}{\begin{tabular}{|p{3cm}|p{5cm}|p{5cm}|}
\hline
\textbf{Classification}       & \textbf{IoV}                                                                    & \textbf{IoAV}                                                                                                         \\ \hline
\textbf{Tracking Behavior}    & Little support for tracking behavior of drivers and passengers         & Behavior monitoring of person, things or data through space and time                                         \\ \hline
\textbf{Situation Awareness}  & Conventional approaches to situational awareness using GPS and cameras & Real-time awareness of physical environment through sensor, automotive radars, and cooperative communication \\ \hline
\textbf{Data Flow}            & Limited, due to specific type of data from vehicles                    & Large, due to connectivity of large number of vehicles and devices                                           \\ \hline
\textbf{Decision Analytics}   & High dependence on humans for analyzing data and making decision       & Assisting human through data visualization and deep analysis                                                 \\ \hline
\textbf{Process Optimization} & Low optimization due to less control of the transportation system      & High optimization due to automated control of self-contained closed system                                   \\ \hline
\textbf{Resource Consumption} & High, due to lack of data and information regarding vehicles           & Low, due to availability of information and efficient utilization of resources across network                \\ \hline
\end{tabular}}
\end{table}

The rest of this paper is organized as follows. In Section II, we discuss the state-of-the-art of IoV. Section III provides the key features and enabling technologies for IoAV while Section IV discusses the architecture of IoAV. Numerical results and their discussion is provided in Section V. Section VI explains some socio-technical challenges of IoAV. Finally, the paper is concluded in Section VII.
  
\section{State-of-the-Art of IoV}

In this section, we discuss the broader concept of IoV and provide some recent studies to highlight the deficiencies of existing IoV model.
\subsection{Introduction to IoV}
Prior to starting a discussion on IoAV, it is worthwhile to describe the broader concept of IoV and highlight its existing deficiencies. There is less doubt that vehicular technology is witnessing a rapid transformation; much similar to the one observed in the domain of sensor networks which ultimately shaped the concept of the Internet of things (IoT). However, unlike IoT, where passive computing devices form a network of intelligent sensors accessible through the Internet, the IoV involves much more mobility and human interventions \cite{7892008}. \textcolor{black}{Human safety and comfort are the indispensable factors in the design of IoV as compared to conventional IoT.} Moreover, there are different levels of communications in IoV including intra-vehicular communications, inter-vehicular or vehicle-to-vehicle (V2V) communications, and vehicle-to-infrastructure (V2I) communications \cite{7892008}. The heterogeneity of communication at different levels is accompanied by the issues of scalability and expanded coverage area for communication. In order to ensure efficient use of resources, vehicles are grouped into clusters with respect to the quality of the wireless channel, and geographical location or mobility of vehicles. These clusters are called the vehicular clouds \cite{gupta2018authorization}, wherein, the data within a vehicular cloud is shared, processed, and frequently uploaded to the Internet cloud for classification of routes and minimizing road accidents.  

\subsection{Recent Advances in IoV}
\textcolor{black}{Recent studies on IoV take into account different aspects ranging from social networking, to load balancing and security.}  An overview of communications technologies for vehicles and taxonomy for IoV was discussed in \cite{fangchun2014overview}. Alam \emph{et al}. in \cite{alam2015toward} provided a review of social IoV networks and discussed different social vehicular applications like ``Road Speak'', ``Social Drive'', and ``Road Sense''. These applications help vehicles on the road to recommend chat groups of similar interest and to exchange road conditions and driving experiences in real-time. They also proposed a message exchange structure using SAE J2735. A content dissemination strategy for cooperative vehicular networks was proposed in \cite{7277118}. The key aim of this study was to minimize the load on cellular traffic using ad-hoc based V2V and Wi-Fi-based V2I communications. To improve the energy efficiency of vehicles, a coalition game was formulated in \cite{kumar2016energy} for a vehicular cloud environment. For each vehicle in the game, the authors proposed demand-supply based payoff mechanism to maximize the efficiency. More recently, the authors of \cite{hagenauer2017vehicular} proposed a micro cloud of vehicles to maintain reliable connectivity between vehicles and infrastructure. In view of the aggregation functionality of the cluster head, the authors showed the need for such a micro cloud. From the security and privacy standpoint, Joy \emph{et al}. in \cite{joy2018internet} discussed various security and privacy challenges for IoV. They also presented different approaches like crowdsourced authentication and haystack privacy for ensuring safe and secure communication in IoV. Using the studies of V2I and V2V communication modes, a predictive strategy for fog nodes was proposed by the Zhang \emph{et al}. in \cite{zhang2017mobile}. In a similar fashion, the authors of \cite{8318667} propose an offloading scheme for real-time management of fog-based IoV traffic. However, it was not clarified how to utilize vehicles beyond the range of fog nodes to offload load for traffic management servers. 

\textcolor{black}{We have observed that recent studies on IoV aim to address issues related to social networking, smart cities, and dissemination/ offloading of content. Very few studies have made an attempt to address the shortcomings of IoV to accommodate autonomous vehicles in the presence of legacy vehicles.} Moreover, since the legacy vehicles are most likely to coexist with the autonomous vehicles in coming decades \cite{litman2017autonomous}, the simplistic design of IoV is expected to under-perform in managing the scale and complexity of communication \cite{gupta2018authorization}. Thus, due to different attributes of legacy and autonomous vehicles, there is a need of infrastructure that is flexible enough to support both types of vehicles while simultaneously making the way for a smooth transition to fully autonomous transportation.   

\section{Features and Enabling Technologies of IoAV}
\textcolor{black}{In this section, we describe the salient features of IoAV and discuss some key enabling technologies for practical realization of IoAV.}

\subsection{Salient Features of IoAV}
In the following, we define the essential features of IoAV to enable the convergence of legacy and autonomous vehicles and to pave the way for a fully autonomous transportation system.
\subsubsection{Organization}
One of the fundamental characteristics of IoAV is management and organization. Here, the intent is to free the human cognition from the perplexing details of vehicle maintenance and operations and to provide a communications platform that performs at peak 24/7. The IoAV would enable vehicles to adjust their operations in the face of software failures, change in communication components, and varying environmental conditions. The IoAV would also help vehicles in monitoring their own use and update softwares when needed. Especially, if the IoAV detects any malicious activity from a vehicle, it should have the capability to isolate it from the network and find a suitable alternative for passengers in the minimum amount of time.   
\subsubsection{Configuration}
The optimal configuration is an important feature of the autonomous system \cite{litman2017autonomous}. For the IoAV, configuring a complex and large-scale system can be error-prone and time-consuming. \textcolor{black}{Thus, the communication system must be able to configure itself in accordance with the high-level policies that clearly specify what is desired by traffic control authorities. It is evident that the IoAV will have to accommodate several third-party components.} Therefore, much like a new cell is seamlessly included in the body, a self-configuring IoAV architecture would be able to integrate these third-party components.
\subsubsection{Optimization}
Tuning the performance of a complex system like IoAV may have an unexpected effect on the behavior of the entire system. \textcolor{black}{Just like a brain optimizes its performance through learning, the IoAV would be able to use a variety of contextual factors, given in Table \ref{tab_2}, to optimize the performance of applications. This new found intelligence would allow the system to recognize the tradeoff between two performance influencing factors.} For example, the system would be able to decide whether to assign downlink resources to an autonomous vehicle that is far away from the roadside unit or to serve two nearby autonomous vehicles using the same amount of transmit power. The IoAV would learn to make appropriate choices by monitoring, experimenting and tuning its own parameters.
\subsubsection{Protection}
This feature of self-protection in IoAV mainly deals with minimizing the effect of cascading failures and provide security against any malicious attacks. The IoAV would be self-sufficient in terms of identifying, localizing and defending both hardware and software against potential attacks. The IoAV can also use previous sensor data and monitor the behavior of autonomous vehicles and, subsequently, anticipate problems to avoid system failures. In case of a failure, the system must be capable to mitigate the effect of cascading failures -- thus avoiding the downfall of the entire vehicular network.

\begin{table}
\centering
\caption{Classification of different contextual factors.}
\label{tab_2}
\begin{tabular}{|p{1.5cm}|p{2cm}|p{8cm}|}
\hline
\textbf{Contextual Factors} & \textbf{Property} & \textbf{Examples}                                                           \\ \hline
\multirow{3}{*}{Technical}  & System            & Load migration ability and security of system                               \\ \cline{2-3} 
                            & Network           & Signal-to-noise ratio (SNR) and channel availability                        \\ \cline{2-3} 
                            & Vehicle           & Power usage and processing capabilities of vehicle                          \\ \hline
\multirow{2}{*}{Usage}      & Goal              & Either enjoyment, exchange of information, or emergency broadcasting        \\ \cline{2-3} 
                            & Attention         & Background (in the form of voice) or foreground (in the form of live video) \\ \hline
\multirow{2}{*}{Social}     & Preferences       & Preferences of users in terms of preferred application                      \\ \cline{2-3} 
                            & Consumption       & Alone or in the form of a group                                             \\ \hline
\multirow{2}{*}{Economic}   & Cost              & Transmission and reception costs                                            \\ \cline{2-3} 
                            & Subscription      & Either service already subscribed/ registered                               \\ \hline
\multirow{3}{*}{Temporal}   & Time of Day       & Services available during a specific period of the day                      \\ \cline{2-3} 
                            & Duration          & Duration for which a service is available                                   \\ \cline{2-3} 
                            & Frequency of Use  & Based on how frequently users prefer a particular application               \\ \hline
\multirow{2}{*}{Physical}   & Velocity          & Speed at which the vehicle is moving                                        \\ \cline{2-3} 
                            & Mobility Pattern  & Urban, sub-urban or highway                                                 \\ \hline
\end{tabular}
\end{table}

\subsection{Key Enabling Technologies}
\textcolor{black}{Tried and tested approaches may not be enough to implement different features of IoAV. However, the remarkable amount of innovation seen in sensor technologies and cloud computing along with some recent developments on the wireless physical layer can provide means to realize the complete benefits of IoAV \cite{jameelinterference}. Motivated by this, the following sub-section provides details of some prospective enabling technologies for IoAV:}
\subsubsection{Intelligent Sensing}
Quite often, it can be observed that the focus of vehicular sensor network is mainly on a single type of sensor network. However, for IoAV, we envision that there would be a need for smart sensor technology consisting of the following three types of sensors:
\begin{enumerate}
\item Detection Sensors
\item Ambient Sensors
\item Backscatter Sensor Tags
\end{enumerate}

\begin{figure*}[!htp]
\centering
\includegraphics[trim={0 0cm 0 0cm},clip,scale=.9]{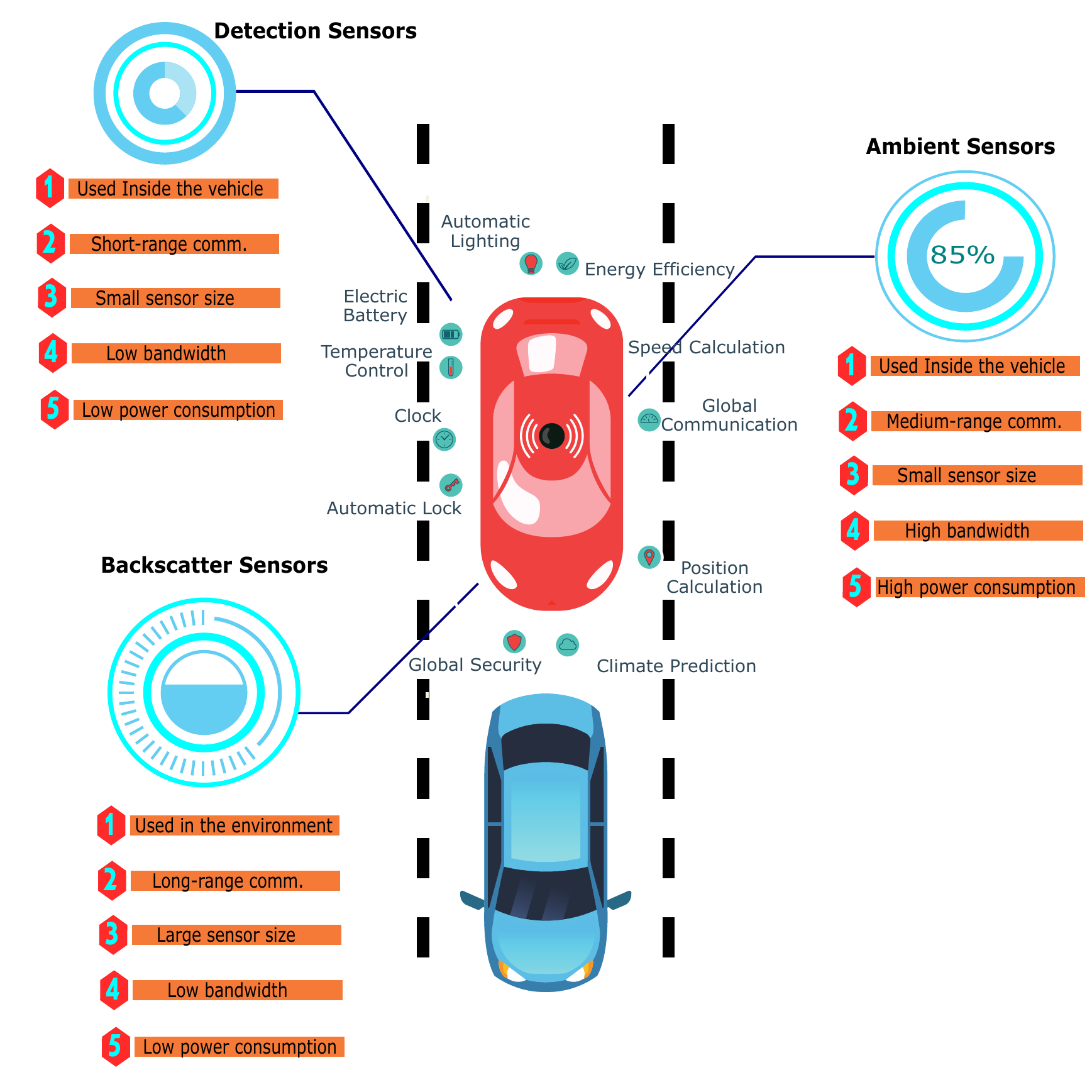}
\caption{\textcolor{black}{Intelligent sensing technology stabilizing the movement of autonomous vehicle in closed spaces. These sensors allow vehicles to navigate by collecting information like speed/ position calculation, battery-life of electric vehicles, and temperature inside/ outside the vehicle. They are also helpful in building a collective perception of the environment by gathering, processing and alayzing real-time information.}}
\label{fig.2}
\end{figure*}

As illustrated in Figure \ref{fig.2}, detection sensors can be mounted in autonomous vehicles to identify different features of the environment and to monitor in-vehicle conditions. Ambient sensors can be part of the environment that can transfer data of legacy and autonomous vehicles to traffic management authorities \cite{litman2017autonomous}. Finally, backscatter sensors can embed connectivity in everyday objects. This can be useful for vehicles to get a collective perception of the outside environment. It can also help in detecting the vitals of the passengers.

\subsubsection{Cloud Computing}
\textcolor{black}{Cloud computing has revolutionized the way we access data on the Internet by virtualizing resources that can be accessed as a service \cite{joy2018internet}.} For IoAV, the concepts of \emph{distribution of integrated resources} and \emph{integration of distributed resources} are expected to be very important. This would also enable encapsulation of distributed resources in the cloud. \textcolor{black}{Recent studies have also indicated the utility of vehicular fog nodes. Vehicular fog nodes use fog computing mechanism to gather and process data from nearby vehicles. Using the croudsourced data over IoAV, autonomous vehicles can perform real-time analytics and optimize the performance of network instead of forwarding all the data to cloud.}

\subsubsection{Vehicular Big Data}
Big data can be used to extract the information behind the collected data from sensors, vehicles and other controlling agents in vehicular networks \cite{zhang2018vehicular}.  Besides four \emph{Vs} of Big Data (i.e., Volume, Velocity, Variety, and Veracity), we propose that vehicular Big Data needs to add three additional \emph{Vs}. These \emph{Vs} are Viability (i.e., the collected information should be viable and valuable), Visibility (i.e., the analysis should be able to reveal invisible information from the data), and Validity (i.e., the duration for which information is considered to be valid). It is worth noting that the first four \emph{Vs} of vehicular Big Data indicate the appearance of data while the last three \emph{Vs} would represent the significance of data.
\subsubsection{\textcolor{black}{Security Techniques}}
\textcolor{black}{Security in vehicular networks is of critical importance as it can be exploited by malicious users to create chaos \cite{zhang2018vehicular}. This issue becomes more complex when there are both legacy and autonomous on the road. To simplify this complexity, physical layer security (PLS) has recently emerged as a complementary solution for conventional cryptographic techniques. Studies have shown that PLS can improve the secrecy performance and friendly jamming techniques can be employed by source, destination or a helper vehicle \cite{jameelinterference}. It is also important to prevent vehicles from broadcasting forged data. Blockchain-based reputation systems, in this regard, can be very helpful to construct a trust model for IoAV. This can be done by implementing efficient certificates and developing a transparent revocation policy.} 
 
\subsubsection{MmWave Vehicular Communications}
The vehicular Big Data and Cloud computing are dependent on the reception of data from sensors and vehicles. Therefore, it is imperative to have Gbps communication links for IoAV using millimeter wave (mmWave). Not only mmWave can be employed for V2V communication, but it can also be used for V2I and intra-vehicle communications. Vehicles can share the raw sensing data from the neighboring vehicles using mmWave V2V links. \textcolor{black}{Other road safety applications can involve mmWave V2I links to gather the sensed data from vehicles and forwarded it to the cloud. Using the high data rate mmWave links, the autonomous vehicles can download real-time maps and live images of the streets.} \textcolor{black}{Several important bands can be used for mmWave-enabled IoAV including some 5G bands (like 28 GHz and 38 GHz) and unlicensed bands (like 60 GHz). As a sub-part of IoAV, automotive radar bands at 24 GH and 76 GHz can also be used for localization of mmWave-based autonomous vehicles.}

\section{Architecture of IoAV}

As with any emerging technology, communication services and systems of IoAV will be characterized by the convergence, assimilation, and integration of different paradigms of communications. \textcolor{black}{In this section, we adopt a bottom-up approach and provide a bird's eye view of the proposed architecture of IoAV. Unlike the architectures proposed for IoV, we find it meritorious to divide IoAV into three layers, i.e., Physical Layer, Virtual Layer, and Management Layer. As illustrated in Figure \ref{fig.3}, each layer has a key contribution in enabling IoAV services while simultaneously allowing construction of a platform for the real-time exchange of information. The physical layer technologies ensure sustainable, low-latency, and high data rate communication for the virtual layer which provides a flexible mechanism to manage resources under the higher-level policies outlined at the management layer. These layers are expected to make IoAV the admixture of both durability and interoperability, diversity and profundity, and most importantly, reliability and flexibility.} The following sub-sections entail a detailed description and contribution of each layer: 

\begin{figure*}[!htp]
\centering
\includegraphics[trim={0 0cm 0 0cm},clip,scale=.4]{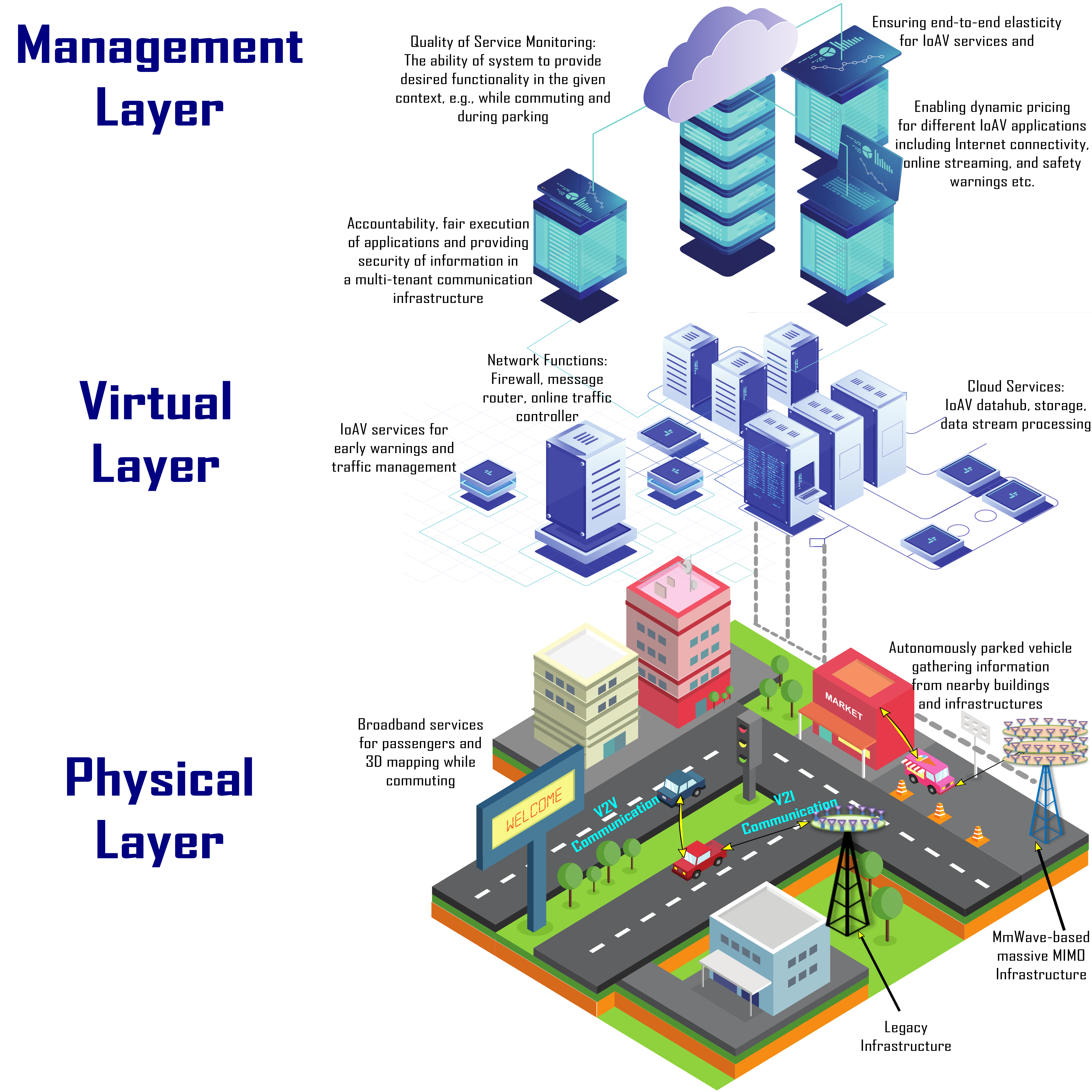}
\caption{Three layers of IoAV. The physical layer is responsible for providing reliable and low-latency broadband services to the vehicles on the road. Virtual layer performs network functions like firewall, message routing, and online traffic controlling. Management layer is tasked to monitor the quality of services while ensuring fair execution of applications and security of information.}
\label{fig.3}
\end{figure*}

\subsection{Physical Layer}

The rapid growth of multimedia services and low-latency communication is influencing the design of future wireless networks. In this regard, quasi- and non-orthogonal multiple access techniques are the two main contenders to meet the future demands of spectrally efficient vehicular communications \cite{li20145g}. Cooperative jamming techniques for vehicular networks have also proven to be very effective for providing protection against eavesdropping attacks. Incentivized cooperation among vehicles can further improve the performance of IoAV without compromising the resources of any individual vehicle. For data rate hungry applications, mmWave communications at 28, 38, and 60 GHz can be exploited to increase the network capacity up to 10 times \cite{li20145g}. Beam tracking and beam alignment technologies using massive multiple-input-multiple-output (MIMO) technology can ensure sensing and detection of autonomous/ legacy vehicles. Wireless power transmission techniques are being developed for charging vehicles in the parking and for cooperative V2V charging of nearby electric vehicles. Above all, the green communications techniques like simultaneous wireless information and power transfer (SWIPT) have the potential to charge small sensors along the road or a highway, thus, paving the way for battery-less charging of sensors \cite{jameelinterference}. 

\subsection{Virtual Layer}
This layer is implemented on top of general purpose hardware that provides a framework for virtualizing the network services. In this way, simple service compositions can be interconnected to create more complex services. Edge and fog computing paradigms can assist virtualization of resources by providing distributed storage and computing infrastructure. Elasticity would be one of the key factors influencing the performance of IoAV. Elasticity techniques studied for centralized Cloud data centers may not be enough for efficiently managing the complex and large-scale communication environment caused by the coexistence of legacy and autonomous vehicles. At peak traffic hours, the demanding infrastructure of IoAV would require end-to-end elasticity, both horizontally and vertically. Along these lines, the virtual layer would enable strong coordination among edge and the Cloud. Given that the Cloud already knows the elasticity demands from the edge, it can provide resources to the vehicles in a more efficient manner. Thus, the virtual layer would have dual responsibilities, i.e., to guarantee elasticity across layers, and to ensure end-to-end elasticity across subsystems like legacy/ autonomous vehicles, network function (edge/ fog systems), and Cloud data centers.

\subsection{Management Layer}

The management layer is the top layer for managing multiple heterogeneous, adaptive, and geo-distributed resources. Besides having the infrastructure- and contextual-awareness \cite{zhang2018vehicular}, efficient utilization of these resources would ensure smooth operation of different computing infrastructure. Especially in the case of autonomous vehicles, this management layer needs to designed to execute continuum set of applications, to expose different view of infrastructure, and to manage Cloud and edge/fog resources. Fair execution of different applications is another task of the management layer by allowing distributed resources to announce availability; an essential to maintain end-to-end elasticity. Issues related to monitoring, accountability, and security (that arise from data distribution, mobility/ interaction of vehicles, multi-tenancy of infrastructure) would also be ensured by the management layer. Stakeholders including transportation control authorities, Internet service providers, and automotive industries would have accessibility to different levels of information. Thus, the management layer would be extremely helpful in creating a disseminated trustworthy communication ecosystem. Finally, the dispersed nature of IoAV would also enable dynamic pricing mechanism through the efficient management of resources. This would help in developing efficient and reliable business models for encouraging and supporting autonomous vehicles.

\begin{figure*}[!htp]
\centering
\begin{tabular}{c}
\includegraphics[trim={0 0cm 0 0cm},clip,scale=.4]{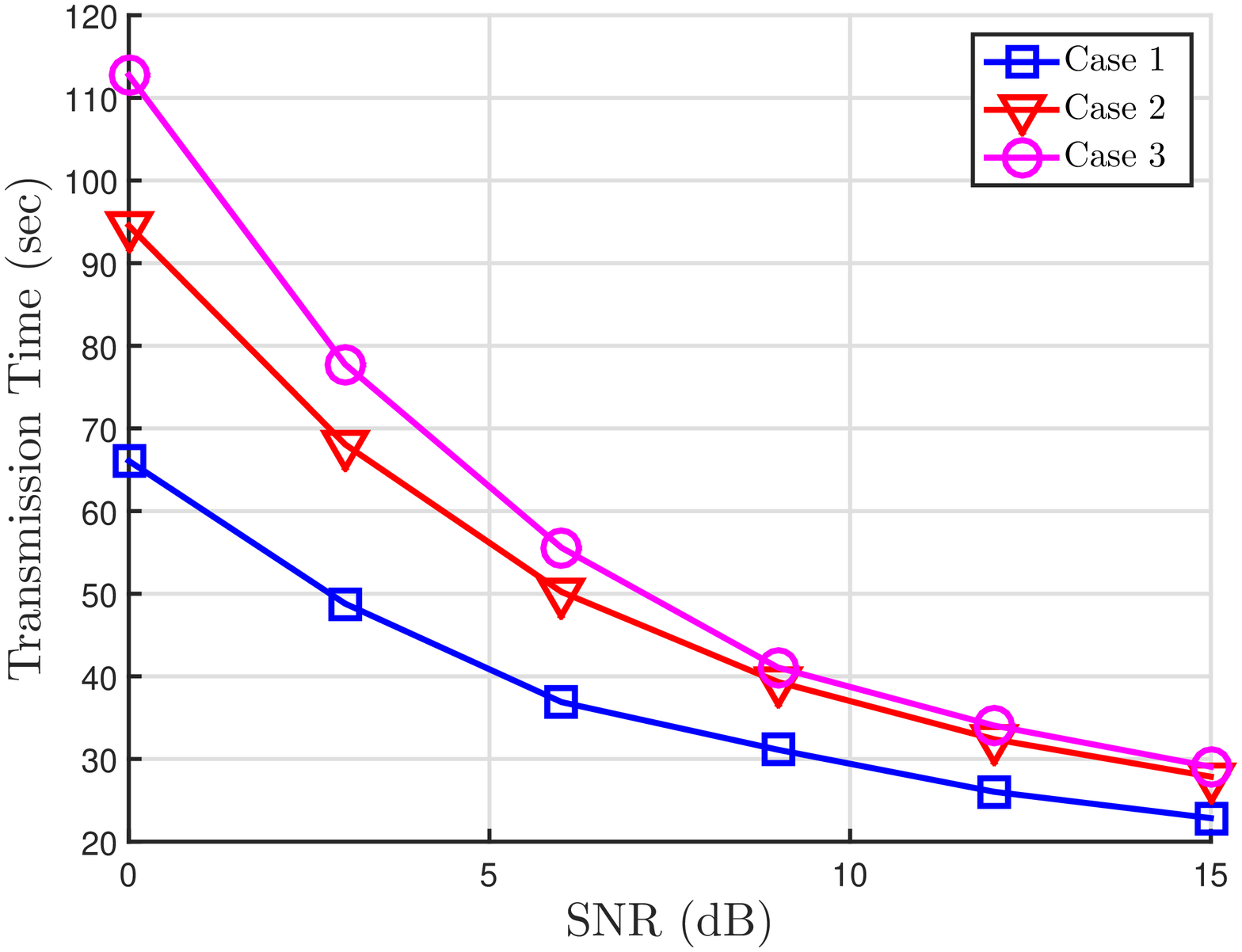}
\\ (a)
\\ \includegraphics[trim={0 0cm 0 0cm},clip,scale=.4]{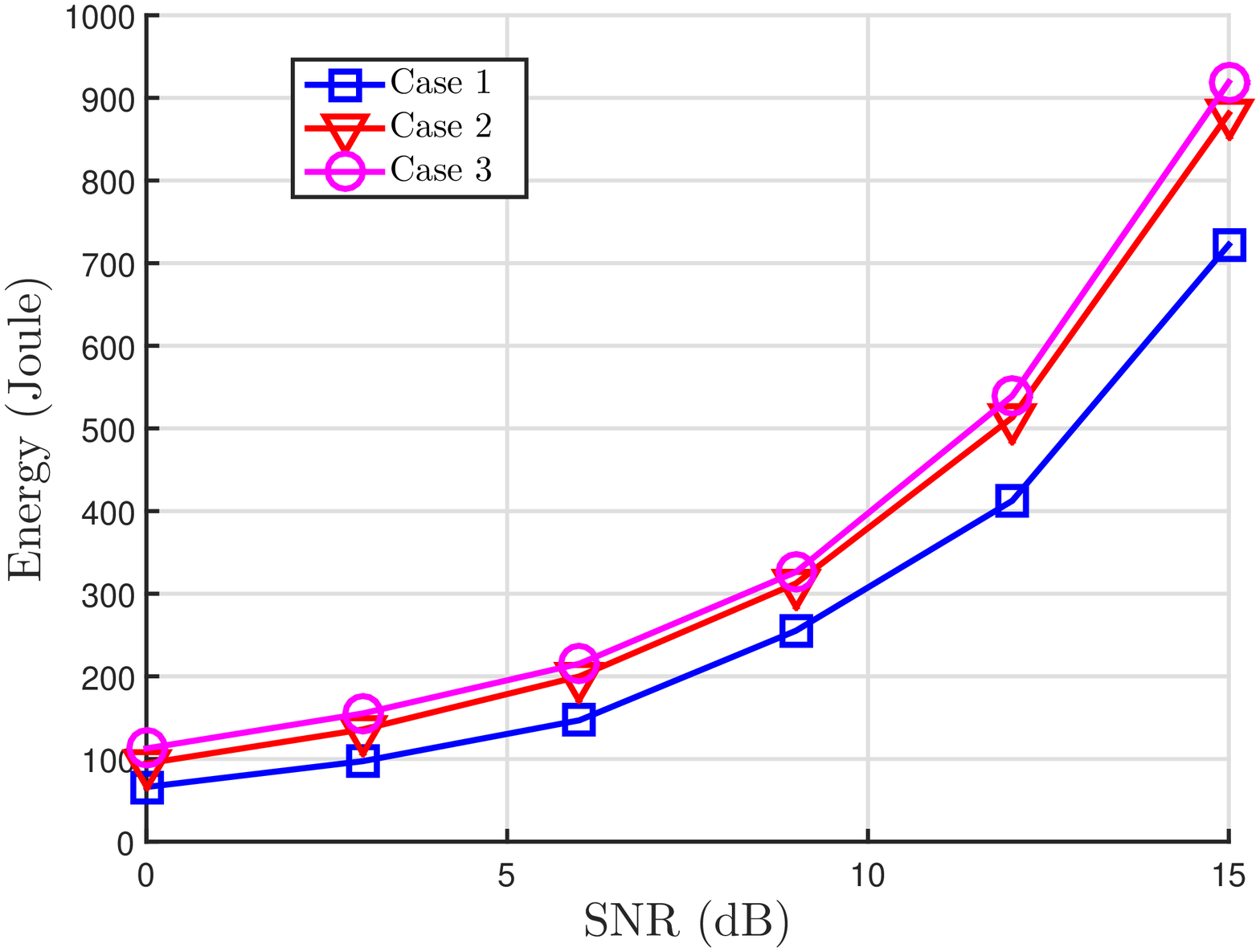}
\\ (b)
\end{tabular}
\caption{\textcolor{black}{Performance of IoAV architecture (a) Transmission time (b) Energy against SNR.}}
\label{fig.res}
\end{figure*}

\textcolor{black}{\section{Performance Evaluation}
This section presents a case study of multi-vehicle offloading scenario to validate the performance of IoAV architecture. Vehicles may be required to transfer the large-volume of data for real-time applications like augmented reality/ virtual reality. Thus, we consider the setup where multiple legacy and autonomous vehicles try to offload their data via wireless access to the resource-rich cloud.  
\subsection{Experimental Setup}
We consider a single-lane highway scenario having both autonomous and legacy vehicles at a uniform distance under Saleh-Valenzuela channel model. \textcolor{black}{We deploy 10 vehicles on the road including 5 autonomous vehicles and 5 legacy vehicles, wherein, each vehicle has 100 MB data to upload to the cloud.} The experimental setup involves the following three cases: 
\begin{itemize}
\item \emph{Case 1:} This case corresponds to the IoAV architecture where mmWave-based massive MIMO architecture is considered to be equipped with 128 antennas and 500 MHz bandwidth. Resource allocation is performed based on the vehicle type. Since autonomous vehicles require permanent connectivity to the Internet, they are considered to use orthogonal multiple access (OMA), whereas, the legacy vehicles use NOMA.
\item \emph{Case 2:} This case is similar to \emph{Case 1}, however, here we consider that both the legacy and autonomous vehicles use orthogonal time-frequency resources. In other words, vehicles in the same beam are considered to use the same orthogonal resources and OMA is performed for all the interfering vehicles. 
\item \emph{Case 3:} In this case, we consider that each MIMO beam only supports a single vehicle and that both types of vehicles use OMA. This also means that the number of RF beams is equal to the number of vehicles in the network.    
\end{itemize}
\subsection{Performance Evaluation}
We consider two performance metrics for offloading the data to the cloud, i.e., transmission time and energy which is consumed during offloading the data. As shown in Figure \ref{fig.res}(a), we have plotted transmission time as a function of signal-to-noise ratio (SNR). It can be seen that the average transmission delay decreases for the \emph{Case 1}, \emph{Case 2}, and \emph{Case 3}. \textcolor{black}{However, \emph{Case 1} significantly outperforms \emph{Case 2} and \emph{Case 3} in terms of offloading delay.} For SNR of just 10 dB, the transmission time reduces 30 sec due to higher data rates of IoAV for offloading the data. To further highlight the utility of IoAV, Figure \ref{fig.res}(b) shows consumed energy against the increasing values of SNR. Note that the energy consumption is lowest for \emph{Case 1}, whereas, it is highest for \emph{Case 3}. The difference in consumed energy between \emph{Case 1} and \emph{Case 3} increases with the increase in SNR which indicates the effectiveness of IoAV at higher values of SNR.}

\section{\textcolor{black}{Social and Technical Challenges of IoAV}}

Due to the involvement of both legacy and autonomous vehicles, the challenges of IoAV range from social to technical domains. In the following, we will describe these challenges in detail.

\subsection{Social Challenges}

Technical progress is rapidly altering the social risk constellation. In many cases, the improvement in technology has raised the level of security and correspondingly shown positive results such as longer life expectancy, and greater prosperity. However, the technical innovation, inevitably, often leads to unintended and unforeseen consequences giving birth to new types of social dilemmas. Therefore, we now enumerate a few social challenges that need to be explored further to gain maximum benefit of IoAV:
\begin{itemize}
\item The dispersion of resources also means dispersion of responsibility. In a technology involving legacy and autonomous vehicles, the system errors cannot be ruled out which can lead to accidents involving property damage or personal injury. Therefore, a clear legal and financial model is required to assign responsibilities to different stakeholders of the IoAV.
\item With the inclusion of a mixed legacy and autonomously systems, the interaction of users with vehicles would completely change. Thus, there will be a need for redefining the boundaries of public and private information. These boundaries can, either positively or negatively, impact the decision-making process of the IoAV in the long run.
\item Closely related to the success of IoAV is modeling the user behaviors of legacy vehicle drivers in the presence of autonomous vehicles. The question is: what qualities do people attribute to autonomous vehicles? Are they valued less competent compared to human drivers, or are they considered to be perfectly functioning vending machines? One of the aims of an introductory IoAV strategy must be painting a realistic picture of autonomous vehicles in the consciousness of all road users.
\item As mentioned above, incentivized cooperation among vehicles can further improve the performance of IoAV. However, the formulation of incentive-based mechanisms in IoAV is a social research challenge. The vehicles may not be willing to share resources like exchanging energy or forwarding data. Efficient incentive (monetary or resource-based) techniques using the behavioral models of drivers can motivate the public to participate in V2V cooperation in IoAV. 
\end{itemize}
\subsection{Technical Challenges}

Technical issues in IoAV may stem from the hardware impairments, incapability of existing cellular networks, and lack of energy management solutions. A brief description of these challenges is provided as follows:
\begin{itemize}
\item Hardware impairments have always been a weak spot of wireless devices. \textcolor{black}{This weakness can become a breaking point for mmWave-enabled IoAV. Due to the involvement of multi-tenant third-party components, inexpensive and un-optimized mmWave communication equipment can hamper the realization of a fully functional IoAV.} In the worst case, it can further deteriorate the performance of network leading to traffic accidents and associated damage to the living things.
\item The existing cellular infrastructure lacks many features that are necessary to implement large-scale IoAV. Although some connectivity arrangements using cellular networks are possible for legacy vehicles, the data rates and latency requirements of autonomous vehicles are very stringent. Moreover, the co-existence of both autonomous and legacy vehicles would require extensive use of machine learning algorithms; something which existing infrastructure is not capable to handle on a large-scale.
\item Most of the components of IoAV will consist of resource-constrained devices, therefore, the lack of suitable energy management techniques can impede the realization of IoAV. It is expected that the future devices (including sensors and backscatter tags) would have large battery-life. \textcolor{black}{Moreover, aside from its obvious benefits, fog computing may result in rapid depletion of batteries of electric vehicles. The vehicular fog nodes are expected to consume more energy as compared to conventional vehicles. Thus, managing real-time information exchange over fog vehicular nodes would be an important yet challenging issue for the IoAV.} 
\item The IoAV will provide services like logistics, intelligent transportation, and vehicle sharing. These services will rely on a range of application requiring low data rates and non-real-time to the more demanding ones having high-speed real-time communications requirements. Therefore, one of the major research challenges for the IoAV would be designing efficient and intelligent routing management techniques \cite{zhang2018vehicular}. We also anticipate that the success of the IoAV will be dependent on smooth interoperability of contemporary and forthcoming mobility management techniques with an aim to achieve a higher number of connected vehicles. 
\end{itemize} 

\section{Conclusion}
The autonomous vehicles are expected to bring radical changes in mobility patterns of the people. However, the current state of technology may not be completely sufficient in realizing the aspiration of seamlessly connected vehicles. We have shown that simply connecting all the vehicles to the Internet is not the solution and, therefore, it is necessary to avoid the pitfalls associated with connecting legacy and autonomous vehicles. In this regard, we have identified deficiencies in the existing IoV model and proposed the IoAV architecture which incorporates both legacy and autonomous vehicles. We have also highlighted the key features, applications, and enabling technologies for IoAV. \textcolor{black}{The initial results on the architecture of IoAV show how the interplay of these enabling technologies would provide performance improvement.} In the end, numerous socio-technical research challenges have also been identified that merit further investigation.
\section*{Acknowledgment}

\ifCLASSOPTIONcaptionsoff
  \newpage
\fi
\bibliographystyle{IEEEtran}
\bibliography{References}
\begin{IEEEbiographynophoto}{Furqan Jameel}
received his BS in Electrical Engineering (under ICT R\&D funded Program) in 2013 from the Lahore Campus of COMSATS Institute of Information Technology (CIIT), Pakistan. In 2017, he received his Master's degree in Electrical Engineering (funded by prestigious Higher Education Commission Scholarship) at the Islamabad Campus of CIIT. In 2018, he visited Simula Research Laboratory, Oslo, Norway. Currently, he is a researcher at the University of Jyv\"askyl\"a, Finland. His research interests include modeling and performance enhancement of vehicular networks, physical layer security, ambient backscatter communications, and wireless power transfer. He was a recipient of the Outstanding Reviewer Award in 2017 from Elsevier.
\end{IEEEbiographynophoto}
\begin{IEEEbiographynophoto}{Zheng Chang}
(S'10 - M'13 - SM'17) received  the Ph.D. degree from the University of Jyv\"askyl\"a, Finland, in 2013.  He was a Visiting Researcher with Tsinghua University, China, in 2013, and the University of Houston, TX, USA, in 2015. He has been honored by the Ulla Tuominen Foundation, the Nokia Foundation, and the Riitta and Jorma J. Takanen Foundation, the Jorma Ollila Grant for his research excellence. He has served as an editor or guest editor for some IEEE journals and magazines, and has received  best paper awards from IEEE TCGCC and APCC in 2017. He is currently an Assistant Professor with University of Jyv\"askyl\"a. His research interests include vehicular networks, green communications, IoT, cloud/edge computing, and security and privacy.
\end{IEEEbiographynophoto}
\begin{IEEEbiographynophoto}{Jun Huang} is with the School of Computer Science, Chongqing University of Posts and Telecommunications, Chongqing, 400065 China.
\end{IEEEbiographynophoto}
\begin{IEEEbiographynophoto}{Tapani Ristaniemi} received the M.Sc. degree in mathematics in 1995, the Ph.Lic. degree in applied mathematics in 1997, and the Ph.D. in wireless communications in 2000 from the University of Jyväskylä, Jyvaskyla, Finland. In 2001, he was appointed as a Professor with the Department of Mathematical Information Technology, University of Jyvaskyla. In 2004, he moved to
the Department of Communications Engineering, Tampere University of Technology, Tampere, Finland,
where he was appointed as a Professor in wireless communications. Prof. Ristaniemi is currently a Consultant and a member of the Board of Directors of Magister Solutions Ltd. He is currently an Editorial Board Member of Wireless Networks and International Journal of Communication Systems.
\end{IEEEbiographynophoto}

\end{document}